# Automatic Analog Beamforming Transceiver for 60 GHz Radios

Shalabh Gupta

University of California, Los Angeles, CA 90095, USA

*Abstract—* We propose a transceiver architecture for automatic beamforming and instantaneous setup of a multi-gigabit-per-second wireless link between two millimeter wave radios. The retro-directive architecture eliminates necessity of slow and complex digital algorithms required for searching and tracking the directions of opposite end radios. Simulations predict <5 micro-seconds setup time for a 2-Gbps bidirectional 60-GHz communication link between two 10-meters apart radios. The radios have 4-element arrayed antennas, and use QPSK modulation with 1.5 GHz analog bandwidth.

## I. INTRODUCTION

The demand for high speed wireless communication systems has made it necessary to move to millimeter wave frequencies. Device scaling in CMOS technologies has enabled low cost communications systems in the unlicensed 57-64 GHz frequency band [1-2]. However, the path loss at these frequencies becomes severe as the effective antenna area scales proportionally with the square of wavelength to maintain omni-directional operation. Thus, use of arrayed antennas providing adaptive beamforming becomes very important.

Millimeter wave frequencies highly favor use of beamforming because of small antenna element sizes and absence of multi-path effects. Adaptive beamforming provides high link gains because of antenna directivity, and the radios can conform to different signal directions. However, adaptive beamforming requires complex digital algorithms for computing the direction of signal arrival and transmission to set up a communication link [3]. Conventional algorithms are very slow, and interface between the digital and analog circuitry also adds significant delays. Even when the radio directions are known, controlling phases of transmit and receive signals to achieve proper directionality becomes difficult. Retro-directive arrays can be used to overcome these limitations [4-7]. In the proposed architecture, we exploit retro-directivity to achieve instantaneous link acquisition, in addition to much simpler analog implementation and automatic phasing of the beamforming arrays.

A retro-directive antenna is an arrayed antenna that transmits in the direction of the incoming electromagnetic signal without prior knowledge of its direction. Retro-directive arrays work on the principle of phase conjugation, as described in details in Section II.

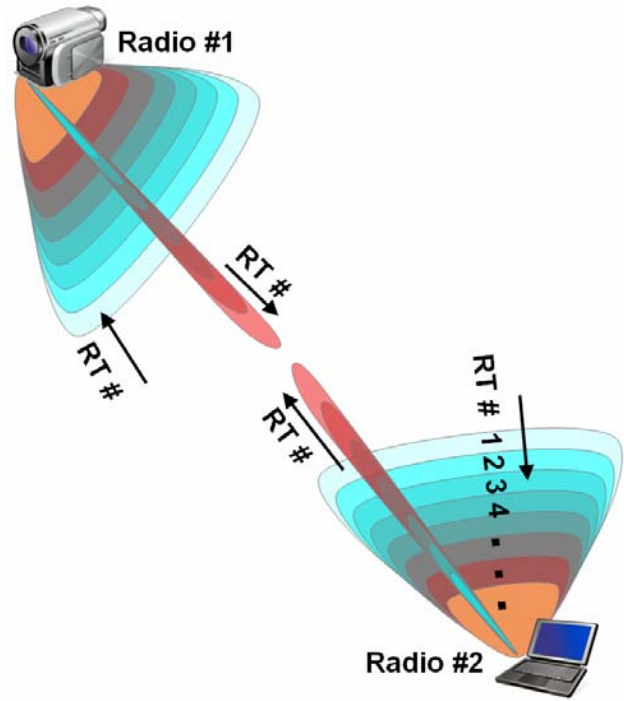

Fig. 1. Link setup in the retro-directive radios. After a few round-trips (RT), the signals from the two radios become directed towards the each other.

Based on the proposed retro-directive radio architecture (described fully in Section III), quick setup of an automatic beamforming RF link is shown in Figure 1. In the absence of an incoming signal, most of the transmitted power goes omni-directionally. When a signal is present at the receiver input, power is transmitted in the direction of the incoming signal. For setup of an RF link, both radios initially transmit omni-directionally (as there are no input signals). Omni-directional transmission is made possible by random phases at different transmit elements. Because of omni-directional transmission, the initial signal power received at the opposite end radios is very weak. However, the transmitters start building directivity towards opposite end radios rapidly in successive transmissions, as they start receiving more and more power after every signal round trip (RT). This positive feedback leads to a quick, automatically tracking RF link between the two radios.

This link setup process is analogous to target acquisition in the retro-directive noise correlation radar [8-9]. However, the threshold power requirements for setting up the link are much smaller as compared to the retro-directive radar since there are active transmitters on both ends.

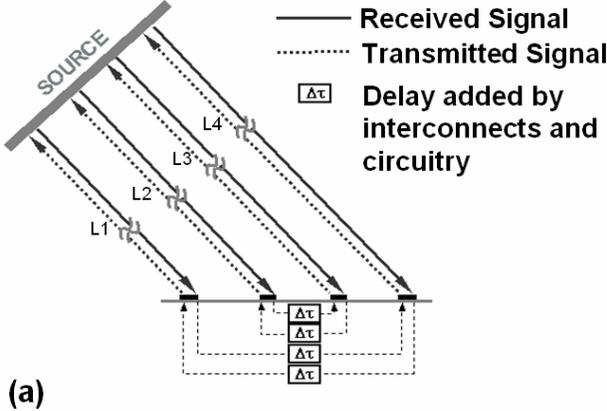

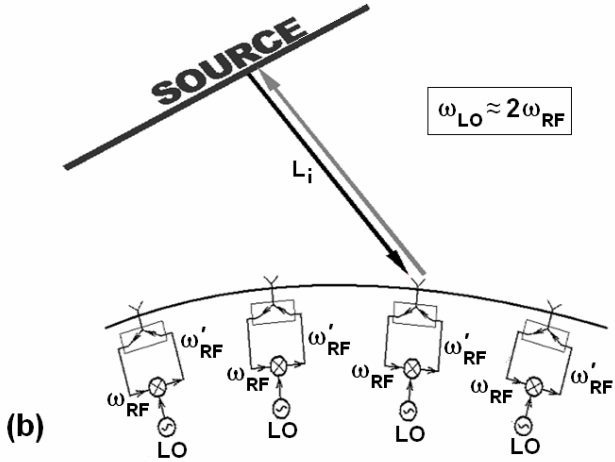

Fig. 2. Retro-directive arrays: (a) Van-Atta array achieves retro-directivity by geometrical positioning of transmit and receive elements. (b) In heterodyne mixing approach, phase conjugation is achieved by mixing RF with LO and using phases of lower sideband IF for transmission.

## II. RETRO-DIRECTIVE ARRAYS

The retro-directive arrays, also known as self-phasing antennas, rely on the principle of phase-conjugation (or phase inversion) for their operation. Traditionally, they have been used as passive transponders, in which any incoming signal is reflected back to the source. When the incoming wave is phase conjugated, both the incoming and the transmitted waves have parallel wave fronts, but travel in opposite directions. Phase conjugation can be achieved by Van Atta array arrangement, or by using heterodyne mixing, as discussed in the following subsections.

### A. Van Atta Array

In Van Atta configuration [4], shown in Fig. 2(a), the signal received at one element is transmitted by another one which is at a "conjugate" position to it, after addition of a constant phase delay due to interconnects and intermediate circuitry. In the figure, for example, signal received at the first element is retransmitted by the fourth element, that received at the second element is transmitted by the third element, and so on. Starting from the source and returning back to the source, the signal undergoes equal phase delays for different paths, as L1 + L4 = L2 + L3, assuming adjacent elements are equidistant to each other. Hence, there is constructive interference of signals transmitted by all elements at the source resulting in directivity towards it. The signal can be transmitted passively or there can be active components in the signal path to modulate and/or amplify the signal before transmission.

### B. Retro-directivity by heterodyne mixing

Use of heterodyne mixing technique to achieve phase conjugation was first proposed in [5]. As in Figure 2(b), when the wave-front from the source is incident at non-zero angle, each array element receives the RF with a different phase due to path length difference. Element $i$ is at distance $L_i$ away from the source and the signal obtained at the $i^{th}$ antenna element becomes $cos(\omega_{RF}t - \beta L_i)$. This signal is mixed with a local oscillator (LO) frequency $\omega_{LO}$ such that $\omega_{LO} > \omega_{RF}$, and low pass filtered to obtain the transmit signal at difference frequency $\omega'_{RF} = \omega_{LO} - \omega_{RF}$, as

$$\cos(\omega_{RF}t - \beta L_i) \times \cos(\omega_{LO}t) \xrightarrow{LPF} \cos(\omega'_{RF}t + \beta L_i) \quad (1)$$

where, $\beta = \omega_{RF}/c$, $c$ = velocity of light.

The resultant signal, when transmitted back by element $i$, has propagation constant $\beta' = \omega'_{RF}/c$, and becomes $cos(\omega'_{RF}t + \beta L_i - \beta' L_i)$ due to the free space propagation delay. When the receive and transmit frequencies are roughly equal, i.e., when $\omega_{LO} \approx 2\omega_{RF}$, the two propagation constants also become roughly same, i.e., $\beta \approx \beta'$. As a result, back at source, the signal phase becomes independent of path differences from different elements, resulting in constructive interference, and hence, directivity towards the source.

To achieve retro-directivity, other frequency plans can be used [10], with the goal being that the excess phases added in the receive paths to the elements are subtracted from the transmit paths for the direction of interest. If the upper side-band signal is used after heterodyne mixing, the transmit array elements should be placed in the same sequence as the receive elements, and if lower sideband is used, the sequence must be reversed. It is important that the receiver and the transmitter frequencies should be close (if inter-element spacing is same for both transmit and receive signals) to avoid any beam pointing inaccuracies. Also, any mismatches in phase delays in different channels must be minimized or calibrated out.

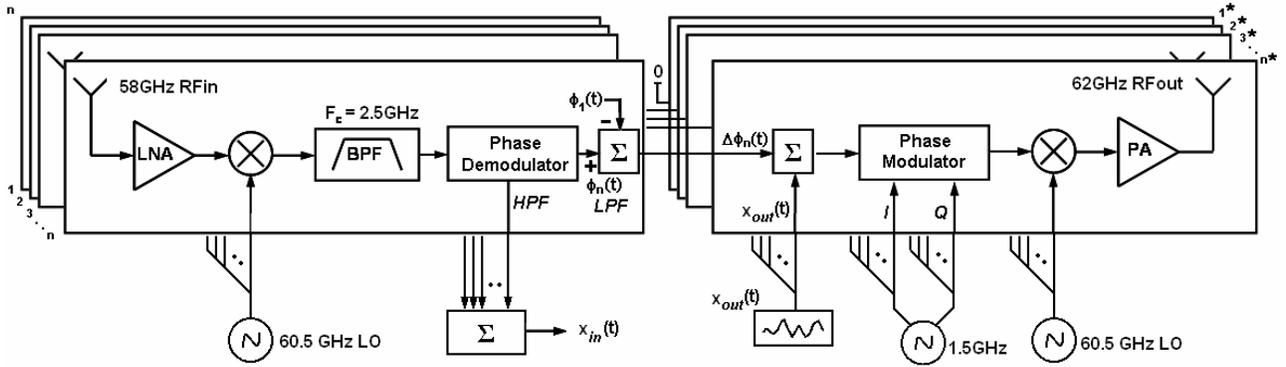

Fig. 3. Proposed 60 GHz radio architecture with retro-directive antennas.

### III. RADIO ARCHITECTURE

Figure 3 shows the architecture of the proposed 60GHz retro-directive radio. As an example, the radio shown in the figure receives signals at 58-GHz and transmits at 62-GHz. On the other hand, the opposite end radio transmits at 58-GHz and receives at 62-GHz. The incoming phase modulated 58GHz RF signal is received by different antenna elements of the radio and is down converted to a convenient 2.5GHz IF frequency in each of the receive chains. After band pass filtering, phase demodulator (PDM) is used to extract the RF phase for each channel. This phase contains both the transmitted data ($x_{in}(t)$) as the high frequency component, and the direction phase $\phi(t)$ as the low frequency component which is used for phase conjugation and retro-directivity. The phase modulating digital data should be encoded such that it doesn't have low frequency components, to avoid interference with the phasing components $\phi_i(t)$. For transmitter beam forming, low pass filtered phase signals are sent to the transmit channels, where the data to be transmitted is added to them. These signals modulate the phase of 1.5GHz IF (intermediate frequency), which is then up converted to 62GHz RF and transmitted by individual antenna elements. Combining the high pass data signals together results in receiver beam forming as the noise components are scaled down due to averaging.

For $i^{th}$ channel phase, instead of using the low pass filtered phase $\phi_i(t)$, $\Delta\phi_i(t)$ is used, where $\Delta\phi_i(t) = \phi_i(t) - \phi_1(t)$, for $i = 1, 2, .. n$. This simple difference operation ensures the there is no drift in the transmit frequency, even if the demodulated input phases are drifting together because of clock offsets between the two radios. This difference operation gives a significant improvement over the retro-directive architecture used in [10], and a lot of instability concerns [11] for a retro-directive link become unimportant.

A key feature of this architecture is that in the absence of an input signal, the power goes omni-directionally since the phases to the transmit elements are uncorrelated. Initial omni-directional transmission is crucial as it provides the cuing signal for opposite end radios to start build up of directivity.

The transmit and receive frequencies are intentionally kept unequal to ensure that there is a significant isolation between the transmitter and the receiver, as both have to operate simultaneously. The frequency plans can be varied a lot depending on availability of the circuit level choices. Different modulation schemes can be used for high bandwidth data modulation, but low bandwidth phase modulation and demodulation are still required for phasing the antennas to achieve directivity. Low pass bandwidth of the phase demodulator dictates the link setup time. Smaller bandwidth results in longer locking times but more stable operation.

### IV. SIMULATIONS AND RESULTS

The radio system discussed in previous section is simulated using Simulink software (from Mathworks), with discrete time simulations and fixed time steps of 5-ps. One of the radio transceivers receives signal at 58-GHz and transmits at 62-GHz. The other radio receives signal at 62-GHz and transmits at 58-GHz. Each of the transceivers has 4 receive antenna elements and 4 transmit elements, aligned linearly at $\lambda/2$ mutual spacing. Both the radios are placed at 42° angles to the broadside direction of the arrays. The path loss is calculated based on a distance of 10 meters which corresponds to a 33-ns propagation delay between the radios (a propagation delay of 35-ns was used in the simulations to account for additional group delay of electronics). For simplicity, the signals are modulated and demodulated directly at the carrier frequencies instead of first converting them to IF.

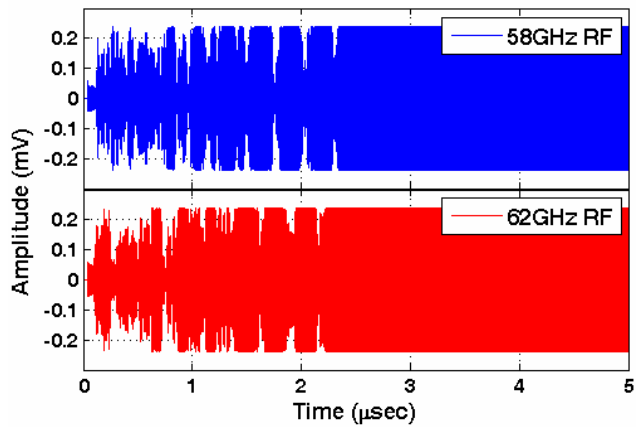

Fig. 4. Signal amplitudes at the input of each receive element as a function of time for the two radios are shown. The stable envelopes beyond 3 micro-seconds indicate that the two radios have locked retro-directively.

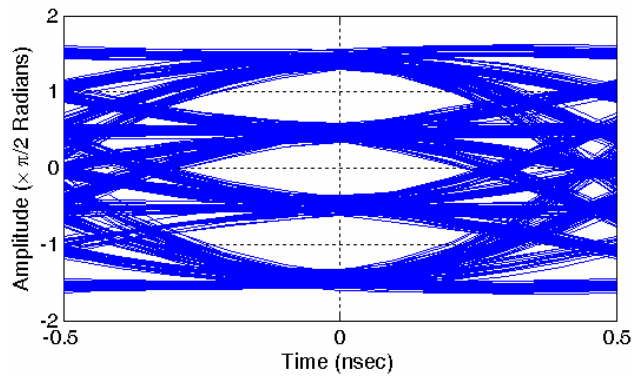

Fig. 5. Eye diagram of the received 2 Gbps QPSK signal obtained after averaging the four phase demodulator channel outputs.

The power transmitted by each antenna element is set as 0-dBm. Since 60-GHz radios assume direct line of sight transmission, use of completely omni-directional antennas is useless. Therefore, the antenna elements are assumed to have a gain of 4-dBi, which limits the operation of the system to a solid angle of about $1.6\pi$ steradians. This gives a path loss of 75-dB for a 60-GHz carrier and element-to-element transmission. The receiver is assumed to have a noise figure of 3-dB and 1.5-GHz band pass FIR filters are used before phase demodulation. Low pass filter used at the phase demodulator output has a 2-MHz bandwidth.

Figure 4 shows the received signal amplitudes at the inputs of antenna elements of the two radios. Because of retro-directivity, the signal strength builds up very quickly (within 3 micro-seconds). The link acquisition times vary depending upon the signal strengths, and at extremely low power levels, the directionality locking between the two radios cannot be achieved. Figure 5 shows the eye diagram of the phase demodulated 2-Gbps data signal. This signal is obtained by combining the demodulator outputs of individual channels and improves the SNR by $\sqrt{N}$, where N is the number of elements. The improvement corresponds to the receiver side beamforming gain.

V. CONCLUSION

The proposed radio architecture can be very useful in instantaneous setup and tracking of multi-gigabit-per-second communication link at millimeter wave frequencies. A 60-GHz 2-Gbps link is demonstrated using computer simulations for two radios at a distance of 10-m and total transmit power of 6-dBm (combined for all 4 array channels) using QPSK modulation. The link acquisition time is found to be 3 microseconds.